\documentclass[twocolumn,showpacs,preprintnumbers,amsmath,amssymb]{revtex4}

\usepackage{graphicx}
\usepackage{dcolumn}
\usepackage{bm}

\begin{document}

\title{Collective resonance modes of Josephson vortices in sandwiched stack of Bi$_{2}$Sr$_{2}$CaCu$_{2}$O$_{8+x}$ intrinsic Josephson
junctions}

\author{Myung-Ho Bae}
\author{ Hu-Jong Lee}%
\affiliation{Department of Physics, Pohang University of Science
and Technology, Pohang 790-784, Republic of Korea}%

\date{\today}

\begin{abstract}
We observed splitting of the low-bias vortex-flow branch in a
dense-Josephson-vortex state into multiple sub-branches in
current-voltage characteristics of intrinsic Josephson junctions
(IJJs) of Bi$_{2}$Sr$_{2}$CaCu$_{2}$O$_{8+x}$ single crystals in
the long-junction limit. Each sub-branch corresponds to a plasma
mode in serially coupled Josephson junctions. Splitting into
low-bias linear sub-branches with a spread in the slopes and the
inter-sub-branch mode-switching character are in good quantitative
agreement with the prediction of the weak but finite
inter-junction capacitive-coupling model incorporated with the
inductive coupling. This suggests the importance of the role of
the capacitive coupling in accurately describing the vortex
dynamics in serially stacked IJJs.

\end{abstract}
\maketitle

Ever since the Josephson coupling was known to be established in
Bi$_2$Sr$_2$CaCu$_2$O$_{8+x}$ (Bi-2212) single crystals along the
adjacent superconducting CuO$_2$ bilayers much research interest
has been focused on the dynamics of Josephson vortices in these
intrinsic Josephson junctions (IJJs) in the long junction limit
\cite{Int,Inductive,Inductive2}. The interest stems mainly from
that the Josephson vortex motion in stacked IJJs provides an ideal
model to study the coupled nonlinear dynamic phenomena and also
provides unique possibilities of applying stacked IJJs to
high-frequency active devices such as sub-millimeter oscillators
and mixers.

In Bi-2212 IJJs, superconducting CuO$_2$ bilayers are much thinner
than the $c$-axis London penetration depth, $\lambda_{ab}$
($\sim$200 nm) \cite{Waldmann}. Then, in a transverse magnetic
field, screening supercurrents are induced in the stacked
superconducting layers, which give rise to the inductive coupling
between adjacent IJJs \cite{Clem}. This coupling is strong enough
that it leads to mutual phase locking of Josephson vortex motion
along the $c$ axis of stacked junctions. It has been predicted
that inductively coupled and thus collectively moving vortices may
resonate with the plasma oscillation modes and induce sub-branch
splitting of current-voltage ($IV$) curves {\it near the bias
voltages corresponding to plasma mode velocities}, which was
confirmed by previous numerical analysis and observations
\cite{Sub}. This phase-locked coherence of the plasma oscillation
may be exploited to develop sub-millimeter oscillator devices.

On the other hand, the thickness of CuO$_2$ bilayers is comparable
to the Debye charge screening length $D$ so that the local charge
neutrality in the bilayers breaks down \cite{Machida2}. This
nonequilibrium charge variation in superconducting bilayers yields
adjacent IJJs are capacitively coupled as well. The purely
capacitive coupling effect was treated theoretically before
\cite{Koyama}. Recently, a model of the capacitive coupling
incorporated with the strong inductive coupling has been proposed,
where the capacitive coupling is taken into account on an equal
footing as the inductive one \cite{Kim}. In spite of the relative
weakness, the capacitive coupling is predicted to lead to
significant changes in the collective vortex dynamics. The model
predicts an $IV$ curve splits into linear multiple sub-branches
with varied slopes, which corresponds to different plasma modes,
{\it in the low-bias region of the collective Josephson vortex
flow}. The spread of the sub-branch slopes is governed by the
strength of the capacitive coupling, represented by a parameter
$\alpha$ [=$\epsilon D^2/(st)$; $\epsilon$ is the dielectric
constant of the insulating layer, $s$ (=0.3 nm) and $t$ (=1.2 nm)
are the thickness of the superconducting and the insulating
layers, respectively], which provides a convenient means to
determine the strength of the capacitive coupling in IJJs.

The dynamics of Josephson vortices in stacked IJJs is known to be
sensitive to the vortex density in the junctions \cite{Koshelev0}.
The collective resonance is usually revealed in a high field range
of $H\gtrsim H_d$ [$=\phi_0/ 2(s+t) \lambda_J$; $\phi_0$ is the
flux quantum and $\lambda_J$ is the Josephson penetration depth].
In this dense-vortex state the non-Josephson-like emission
\cite{Hechtfischer} and the Shapiro resonance steps
\cite{Latyshev2} have been observed, confirming the coherence of
the Josephson-vortex-lattice motion over the whole stacked IJJs.
Especially, for a higher field range of $H\gtrsim H_{hd}$
(=$\phi_{0}/ (s+t) \lambda_{J}$), the transverse inter-vortex
spacing becomes close to $\lambda_{J}$ \cite{Bulaevskii}. In this
range the Josephson current along the length of a junction
distributes almost sinusoidally so that the collective resonance
is expected to be stronger. Little experimental studies, however,
have been done for the collective motion in this highly
dense-vortex region.

In this study, we investigated details of multiple plasma modes
manifested in the Josephson vortex-flow branch (JVFB) of stacks of
IJJs in a high magnetic field above $H_d$. In all the three
samples examined sub-branch splitting was seen to start occurring
for a field exceeding $H_d$. Especially, in the JVFB for a field
above $H_{hd}$, we obtained clear hysteretic multiple sub-branches
and mode switching between them. The linear sub-branches for
different collective vortex modes exhibited a spread in the
slopes. The values of the capacitive coupling parameter $\alpha$,
estimated from the spread, for the three samples under study were
within the range of the theoretical prediction of the
hybrid-coupling model \cite{Kim}. In addition, mode velocities
estimated from the inter-branch switching voltages were in
reasonable quantitative agreement with the prediction of the
model.

Slightly overdoped Bi-2212 single crystals used in this study were
grown by the solid-state-reaction method \cite{Nam Kim}. Samples
were micropatterned on the crystals into the geometry of a stack
of junctions sandwiched between top (400 nm thick) and bottom (100
nm thick) Au electrodes [see the inset of Fig. 1(a)]. The sizes of
the three stacks were 17$\times$1.5 $\mu$m$^2$ (SS1),
16$\times$1.5 $\mu$m$^2$ (SS2), and 16$\times$1.4 $\mu$m$^2$
(SS3). Since each superconducting bilayer in a stack of IJJs is
much thinner than $\lambda_{ab}$, Josephson vortices in a usual
mesa structure \cite{Doh} are supposed to be strongly coupled to
those in the basal stack underneath the mesa, which may
significantly distort the dynamics of the Josephson vortices in
the mesa itself. In this study the basal stack was removed by
using the double-side-cleaving technique \cite{Wang} to overcome
this problem. The detailed sample fabrication process is described
elsewhere \cite{Bae}. Eliminating the basal part turned out to be
essential to obtain ideal vortex-flow characteristics of stacked
IJJs. This sandwiched-stack geometry allowed to examine the
collective vortex dynamics over the whole thickness of a stack.
The magnetic field was aligned in parallel with the plane of
junctions within the resolution of 0.01 degree to minimize the
pinning of Josephson vortices by the formation of pancake vortices
in CuO$_2$ bilayers \cite{Hechtfischer2}. All the measurements
were done at 4.2 K in a two-terminal configuration [Fig. 1(a)].
The combined contact resistance for bottom and top interfaces
($\sim$2 k$\Omega$ for SS1 and SS2, and $\sim$600 $\Omega$ for
SS3) was subtracted numerically.
\begin{figure}[t]
\begin{center}
\leavevmode
\includegraphics[width=0.78\linewidth]{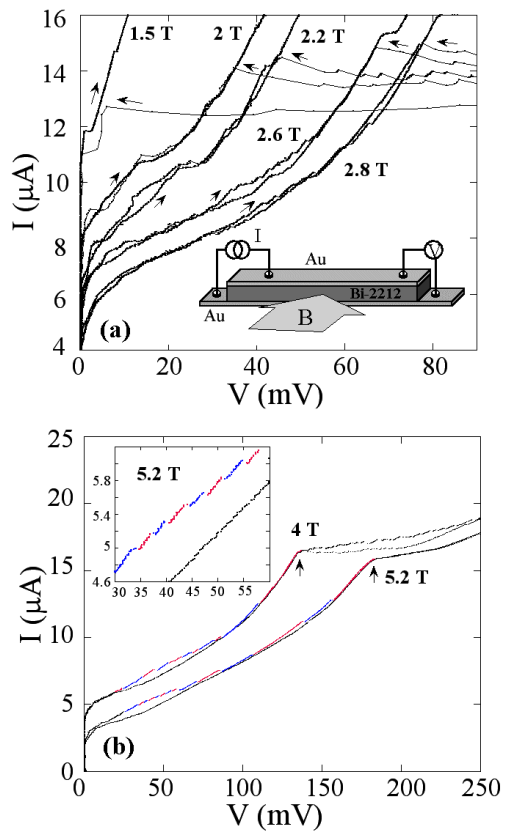}
\caption{(color). Josephson vortex-flow branches of SS1 in field
ranges of (a) $H \sim H_d$ and (b) $H \sim H_{hd}$ (the 4-T curve
is shifted upward by 1 $\mu$A). Neighboring sub-branches are
colored alternately for clarity. The arrows in (a) indicate the
current sweep direction. The arrows in (b) indicate the boundaries
between the vortex-flow and the quasiparicle regions, defining the
parameter $V_c^{vf}$. Inset of (a): the measurement configuration.
Inset of (b): low-bias vortex-flow region at 5.2 T}
\end{center}
\end{figure}
\begin{table}
\caption{\label{tab:tableI}Sample parameters. $N$ is the total
number of IJJs, $J_c$ the critical tunneling current, $\beta_c$
[=$(4I_c/\pi I_r)^2$] the McCumber parameter, $\lambda_{J}$ the
Josephson penetration depth, and $\alpha$ the capacitive-coupling
parameter.}
\begin{ruledtabular}
\begin{tabular}{ccccccc}
Sample&stack size &$N$&$J_c$&$\beta_c$&
$\lambda_J$&$\alpha$\\
 number& ($\mu$m$^2$)& & (A/cm$^2$) & &($\mu$m)  \\
\hline
SS1 &17$\times$1.5 &55&1000 &2000 &0.31 &0.35\\
SS2 &16$\times$1.5 &25&1500 &1300 &0.25 &0.17\\
SS3 &16$\times$1.4 &60&1270 &2200 &0.27 &0.24\\
\end{tabular}
\end{ruledtabular}
\end{table}

The sample parameters are listed in Table I. The number of IJJs in
a sandwiched stack, $N$, was estimated from the number of
zero-field quasiparticle curves (not shown). The McCumber
parameter $\beta_{c}$ was obtained from the zero-field return
current $I_r$ [as denoted by arrows in Fig. 1(a) for decreasing
bias]. Fig. 1 shows the JVFB of sample SS1 with a single up-down
bias sweep. The values of $H_{d}$ and $H_{hd}$ for SS1 estimated
with $\lambda_{J}$ were 2.2 and 4.4 T, respectively. Thus, the
data in Figs. 1(a) and 1(b) correspond approximately to the two
dense-vortex ranges, $H\sim H_d$ and $H\sim H_{hd}$, respectively.
The average slope, $\Delta V/\Delta I$, of the JVFB keeps
increasing in proportion with the magnetic field intensity. A few
voltage jumps are seen in the JVFB for fields beyond $\sim$2.6 T
in the bias range above $\sim$40 mV. This unusual feature, taking
place in the vortex-flow state for a bias below the
quasiparticle-state return current $I_r$ in each field, is caused
by switching of coherent vortex-lattice motion between different
plasma modes. Coherently moving vortex lattice remains in a
certain transverse plasma mode as long as the vortex-lattice
velocity is smaller than the propagation mode velocity of the
plasma oscillation. Exceeding the mode velocity for a higher bias
current, however, the resonating dynamic state of the Josephson
vortex lattice suddenly switches to the adjacent plasma mode with
a higher propagation velocity, giving rise to a voltage jump
observed. The onset field of the observed voltage jumps almost
coincides with the value of $H_d$.

Clearer voltage jumps in the JVFB are seen for a higher vortex
density in the field range of $H\gtrsim H_{hd}$ as in Fig. 1(b),
as the resonant coupling between the vortex-lattice motion and the
plasma oscillation is strengthened. In this field range the
hysteresis in the quasi-particle branches for $V>V_c^{vf}$
(denoted by the arrows) in Fig. 1(b) is almost completely
suppressed while the voltage jumps become more evident in JVFB.
The inset of Fig. 1(b) shows the distinct voltage jumps in the
lower bias region for 5.2 T. One notes that, in Fig. 1(b), all the
voltage jumps take place in the low-bias flux-flow region bounded
by $V_c^{vf}$. The voltage interval between the neighboring jumps
tends to increase for higher biases, which is consistent with the
prediction of both the purely inductive and the
inductive-capacitive hybrid coupling models.

Fig. 2(a) displays the JVFB of the sample SS2 (with $N$=25) taken
for $H$=5.8 T. For the measurement the bias current was repeatedly
swept up and down so that detailed hysteretic feature of the
sub-branches were traced out. These multiple-sweep data clearly
indicate that the voltage jumps observed in the single-sweep data
as seen in Fig. 1(b) were indeed caused by the inter-sub-branch
mode switching. Since the $H_{hd}$ for this sample is 5.4 $T$ the
data in Fig. 2 correspond to the highly dense vortex state of
$H\gtrsim H_{hd}$, expectedly with many resonant vortex states as
in the sample SS1 in the similar field range. About 22 multiple
sub-branches were obtained in SS2, which was close to the number
of the IJJs $N$=25 in the stack. Details of the vortex-flow region
of 5.2 T curve of SS1, as shown in Fig. 1(b), are also illustrated
in Fig. 2(b) for a quantitative analysis.

In Figs. 2(a) and 2(b), all the discernible sub-branches exhibit
linear $IV$ characteristics in the low-bias vortex-flow region for
$V<V_c^{vf}$, the occurrence of which is in agreement with the
prediction of the inductive-and-capacitive hybrid coupling model.
By contrast, the purely inductive coupling yields sub-branch
splitting near the voltages corresponding to the plasma mode
velocities. As denoted by the grey region in Fig. 2 the linear
sub-branches, when extrapolated to the low-bias region, converges
to a single intercept point with a positive value $I_p$ on the
current axis. It confirms the vortex-flow nature of the
sub-branches with finite vortex pinning, that may have been caused
by any defects in the IJJs or by any pancake vortices formed in
CuO$_2$ layers by a misaligned field component. The pinning
current $I_p$ was 2.07 $\mu$A and 0.64 $\mu$A in SS1 and SS2,
respectively. For an analysis of the finite spread in the slopes
of the linear sub-branches we adopt fitting to the hybrid coupling
model \cite{Kim}, while incorporating the effect of the finite
vortex pinning (represented by $I_p$) and the change of the vortex
density with applied magnetic fields ($H_{ext}$) as
\begin{equation}
V=N(I-I_p)[\frac{R_n}{A_n}+\frac{kH_{ext}}{I_c}].
\end{equation}
Here, $A_n$=$1+2\alpha[1-\mbox{cos}(n\pi/(N+1))]$, $I_c$ is the
critical Josephson current at a given applied field, and $k$ is a
constant. Since samples are in the quasiparticle state for
$V>V_c^{vf}$ we define the current corresponding to $V_c^{vf}$ in
a given field as the critical current $I_{c}$ in Eq. (1). Also,
the vortex velocity for $I>I_p$ is assumed to increase linearly
with the bias current.

\begin{figure}
\begin{center}
\leavevmode
\includegraphics[width=0.74\linewidth]{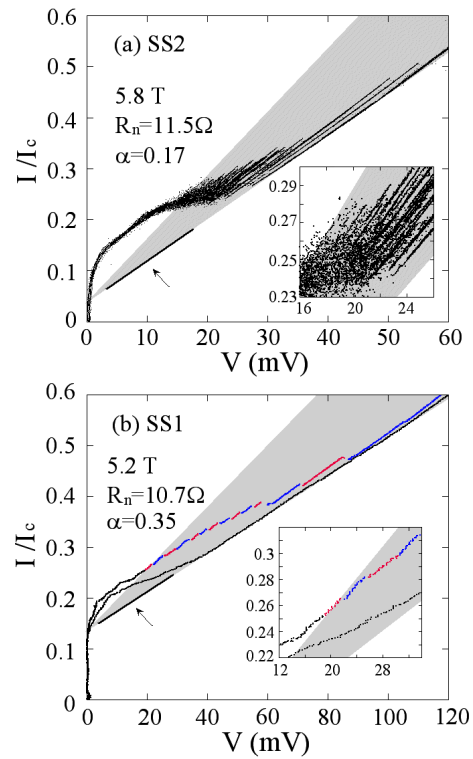}
\caption{(color). Josephson vortex-flow sub-branches (a) for SS2
at 5.8 T and (b) for SS1 at 5.2 T with the spread of linear
sub-branches denoted by the gray region. Insets: details of the
sub-branches in the low-bias range.}
\end{center}
\end{figure}

Without the capacitive-coupling effect ($\alpha$=0) the low-bias
sub-branches with spreaded slopes, as exhibited by the grey
regions in Figs. 2(a) and 2(b) for SS1 and SS2, respectively, are
supposed to reduce to single lines (pointed by arrows), where the
value of $k$ determines the slope of the lines. Increasing
$\alpha$ the spread of sub-branch slopes increases from the
$\alpha$=0 lines to the left. In the very low bias region the JVFB
in both samples becomes chaotic.

The parameter $\alpha$ was determined from the best fit of the
spread to Eq. (1), where the normal-state resistance $R_n$ was
fixed separately following the fitting procedure in Ref.
\cite{Won}. The best-fit values of $\alpha$ for SS1 at 5.2 T, SS2
at 5.8 T, and SS3 at 4.15 T were 0.35, 0.17, and 0.24,
respectively. These values are in the range of theoretical
estimate of the hybrid coupling model \cite{Kim}, $0.1 \sim 0.4$,
and in reasonable agreement with the recently observed values of
$0.36 \sim 0.44$ in SmLa$_{1-x}$Sr$_x$CuO$_{4-\delta}$
\cite{Helm}. The corresponding values of the $k$ for SS1, SS2, and
SS3 were 0.32, 0.14, and 0.38 mV/T, respectively.

\begin{figure}[t]
\begin{center}
\leavevmode
\includegraphics[width=0.7\linewidth]{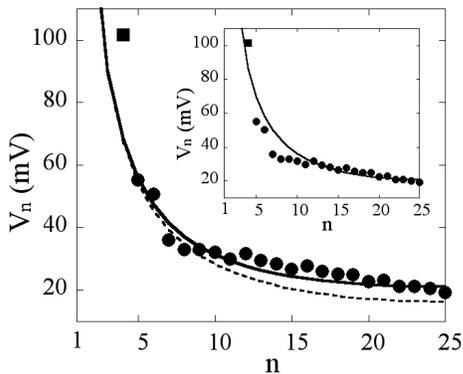}
\caption{Mode-switching voltages (filled circles) and $V_c^{vf}$
(filled square) as a function of mode index $n$ for SS2. The solid
line is the best fit to the hybrid coupling model. The dotted line
is obtained for the inductive coupling only, with the same
parameters as used for the solid line. Inset: the best fit to the
purely inductive coupling model.}
\end{center}
\end{figure}

Fig. 3 displays the mode-switching voltages (filled circles) and
boundary voltage $V_c^{vf}$ (filled square) for SS2 as a function
of the mode index $n$. Out of 25 expected modes corresponding to
$N$ the three lowest-index modes were not observed, however,
presumably because they were located beyond $V_c^{vf}$. The solid
line is the trace calculated with $\alpha$=0.17 and the best fit
value for the Swihart velocity $c_0=$1.05$\times$10$^5$ m/s. The
resonance voltage of the $n$th mode is obtained from the relation
\cite{Hechtfischer} $V_n=Nc_n H_{ext}(t+s)$. The mode velocity of
the transverse plasma oscillation ($c_n$) is expected to be
enhanced by a factor of $\sqrt{A_n }$ from that ($c_n^0$) of the
purely inductive coupling mdoel ($A_n =1$) by the nonvanishing
capacitive-coupling effect \cite{Kim} as in the relation
\begin{equation}
c_n=c_0\sqrt{\frac{A_{n}}{1-\mbox{cos}(\pi
n/(N+1))}}=c_n^0\sqrt{A_n}\ ,
\end{equation}
where the mode index $n$ runs from 1 to $N$. The dotted line
corresponds to the case of purely inductive coupling ($\alpha$=0)
with the same Swihart velocity as used for the solid line. One
sees that the enhancement of the mode velocities due to
nonvanishing capacitive coupling becomes more evident for high
mode indices, giving a maximum enhancement for $c_N$ as
$c_N=c_N^0[(1+4\alpha)]^{1/2}$. The inset of Fig. 3 illustrates
the best fit of the mode voltages to the purely inductive-coupling
model, with the best-fit value of $c_0$ to be 1.35$\times$10$^5$
m/s. In this case, mode voltages for low-index modes are seen to
fall significantly below the model expectation. This trend was
also observed previously in the motion of microwave-induced
Josephson vortices, although the vortex density was much lower
than in this study \cite{Doh}.

The Swihart velocity for SS2, estimated using the resistively
shunted junction model, was $c_0=2\pi f \lambda_J$=8.7$\times$
10$^4$ m/s. In the calculation we used the Josephson plasma
frequency $f_{pl}$ (=$\sqrt{I_c/2\pi \Phi_{0}C_J}$) of 55 GHz,
estimated from the critical current $I_c$=0.358 mA and the
junction capacitance $C_J$=9 pF determined from $\beta_c$ and
$R_n$. The value of $c_0$ is close to the one obtained for the
hybrid-coupling model but tends to deviate more from the
predication of the inductive-coupling model.

In summary, the observed characteristics of the sub-branch
splitting in the vortex-flow region were in reasonable
quantitative agreement with the inductive coupling incorporated
with the capacitive coupling. The multiple sub-branches in JVFB
were shown to be caused by the inter-junction capacitive coupling
and the value of the coupling parameter $\alpha$ estimated from
the spread of multiple branches was in good agreement with the
theoretical prediction of the hybrid coupling as well as the
previous observation in optical measurement \cite{Helm}. The
quantitative identification of the multiple sub-branches in this
study confirmed the significance of the capacitive coupling in
addition to the inductive coupling in accurately describing the
dynamics of Josephson vortices in serially stacked IJJs. For
further detailed analysis of the Josephson vortex dynamics
consideration of the charge-hole imbalance coupling \cite{Ryndyk1}
and the dissipation effect of quasiparticles in the $ab$
plane\cite{Koshelev} may be required.

We thank Ju H. Kim for valuable discussions. This work was
supported by the National Research Laboratory project
administrated by KISTEP.


\begin{thebibliography}{00}
\bibitem{Int} R. Kleiner {\it et al.}, Phys. Rev. Lett. {\bf 68}, 2394 (1992);
                  A. Yurgens {\it et al.}, Phys. Rev. B {\bf 53}, R8887 (1994).
\bibitem{Inductive} S. Sakai {\it et al.}, J. Appl. Phys {\bf 73}, 2411 (1993);
                   R. Kleiner, Phys. Rev. B. {\bf 50}, 6919 (1994);
                   M. Machida {\it et al.}, Physica(Amsterdan) {\bf 330C}, 85 (2000).
\bibitem{Inductive2}J. U. Lee {\it et al.}, Appl. Phys. Lett. {\bf 67}, 1471 (1995);
                   V. M. Krasnov {\it et al.}, Phys. Rev. B {\bf 61}, 766 (2000).
\bibitem{Waldmann} O.Waldmann {\it et al.}, Phys. Rev. B 53, 11825 (1996).
\bibitem{Clem} J. R. Clem and M. W. Coffey, Phys. Rev. B {\bf 42}, 6209 (1990).
\bibitem{Sub} R. Kleiner {\it et al.}, Phys. Rev. B. {\bf 50}, 3942 (1994);
              A. Irie {\it et al.}, Appl. Phys. Lett. {\bf 72}, 2159 (1998);
              R. Kleiner {\it et al.}, Physica (Amsterdam) {\bf 362C}, 29 (2000).
\bibitem{Machida2} M. Machida {\it et al.}, Phys. Rev. Lett. {\bf 83}, 4618 (1999).
\bibitem{Koyama} T. Koyama and M. Tachiki, Phys. Rev. B. {\bf 54}, 16183 (1996);
\bibitem{Kim} J. H. Kim, Phys. Rev. B {\bf 65}, 100509 (2002);
              J. H. Kim and J. Pokharel, Physica C {\bf 384}, 425 (2003).
\bibitem{Koshelev0} A. E. Koshelev, Phys. Rev. B {\bf 66}, 224514 (2002).
\bibitem{Hechtfischer} G. Hechtfischer {\it et al.}, Phys. Rev. Lett. {\bf 79}, 1365 (1997).
\bibitem{Latyshev2} Yu. I. Latyshev {\it et al.}, Phys. Rev. Lett. {\bf 87}, 247007 (2001).
\bibitem{Bulaevskii} L. Bulaevskii and J. R. Clem, Phys. Rev. B {\bf 44}, 10234 (1991).
\bibitem{Nam Kim} N. Kim {\it et al.}, Phys. Rev. B. {\bf 59}, 14639 (1999).
\bibitem{Doh} Y.-J. Doh {\it et al.}, Phys. Rev. B. {\bf 63}, 144523 (2001).
\bibitem{Wang} H. B. Wang {\it et al.}, Phys. Rev. Lett. {\bf 87}, 107002 (2001).
\bibitem{Bae} M. H. Bae {\it et al.}, Appl. Phys. Lett. {\bf 83}, 2187 (2003).
\bibitem{Hechtfischer2}G. Hechtfischer {\it et al.}, Phys. Rev. B {\bf 55}, 14638 (1997); A. Irie {\it et al.}, Phys. Rev. B {\bf 62}, 6681 (2000).
\bibitem{Won} H. Won and K. Maki, Phys. Rev. B {\bf 49}, 1397 (1994).
\bibitem{Helm} Ch. Helm {\it et al.}, Phys. Rev. Lett. {\bf 89}, 57003 (2002).
\bibitem{Ryndyk1} D. A. Ryndyk {\it et al.}, Phys. Rev. B {\bf 64}, 052508 (2001).
\bibitem{Koshelev} A. E. Koshelev and I. S. Aranson, Phys. Rev. Lett. {\bf 85}, 3938 (2000).
\end{thebibliography}
\end{document}